\documentclass[mathleft
]{an}
\usepackage{graphicx}
\usepackage{times}

\begin{document}

% The following seven commands are intended for editorial usage and should be ignored by
% the author(s).
\Pagespan{789}{}% Document's page range.
% If second parameter is left empty, the last page is computed automatically.
\Yearpublication{0000}%
\Yearsubmission{0000}%
\Month{00}%
\Volume{000}%
\Issue{00}%
% \DOI{This.is/not.aDOI}%

\title{Kinematics of nearby OB3 stars \\with interstellar CaII line distances}

\author{V.V. Bobylev\inst{1,2}\fnmsep\thanks{Corresponding author:
  \email{vbobylev@gao.spb.ru}\newline}
%Example
%for footnote, note the usage of the \texttt{fnmsep}
%command as separator between institute number and footnote mark}
\and  A.T. Bajkova\inst{1} }
\titlerunning{Kinematics of nearest OB3 stars with interstellar CaII line distances}
\authorrunning{V.V. Bobylev \& A.T. Bajkova}
\institute{Central (Pulkovo) Astronomical Observatory, Pulkovskoye
Shosse 65/1, St.-Petersburg, 196140, Russia \and Sobolev
Astronomical Institute, St. Petersburg State University,
Universitetskii pr. 28, Petrodvorets, 198504, Russia }

\received{00 Mon 0000} \accepted{00 Mon 0000} \publonline{later}


\abstract{%
We tested the distances derived from the equivalent widths of
interstellar CaII spectral lines by Megier et al. 2009. To this
end, we used a sample of nearby 126 young OB3 stars ($r<1$ kpc)
with known proper motions and line-of-site velocities. It is shown
that these stars are tightly bounded with the Gould Belt
structure. Most part of this sample (about 100 stars) show the
same kinematics as the sample of distant OB3 stars. Their
galactocentric radial velocities are in good agreement with the
following spiral density wave parameters: amplitude of radial
perturbations $f_R\approx$12~km/s, wavelength
$\lambda\approx2.3$~kpc and phase of the Sun in spiral wave
$\chi_\odot\approx-90^\circ$. But we revealed 20 stars with
absolutely unusual kinematical features. Their galactocentric
radial velocities show a wave, biased on $\approx180^\circ$ with
respect to the wave, found from the whole sample. The idea of
superposition of two spiral patterns seems to be probable.
 }

\maketitle

\section{Introduction}
For study of Galactic kinematics various data are used. These are
line-of-site velocities of HI and HII clouds with distances
obtained using terminal-velocity procedure (Clemens 1985;
McClure-Griffiths~\&~Dickey 2007;  Levine, Heiles \& Blitz 2008);
cepheids with a distance scale based on a period-luminosity
relation, open star clusters and OB associations with photometric
distances (Mishurov~\&~Zenina 1999; Mel'nik, Dambis \& Rastorguev
2001; Zabolotskikh, Rastorguev \& Dambis 2002; Bobylev, Bajkova \&
Stepanishchev  2008; Mel'nik~\&~Dambis 2009); masers with
high-precision VLBI trigonometric parallaxes (Reid et al.~2009;
McMillan~\&~Binney 2010; Bobylev~\&~Bajkova 2010;
Bajkova~\&~Bobylev 2012).

The most young massive stars of high luminosity (OB stars) are of
great interest for the kinematics connected with Galactic spiral
density wave study, because these stars were not able to withdraw
from their birth places during their life time and therefore are
good tracers of Galactic spiral structure.

The recent work by Bobylev~\&~Bajkova (2011) was devoted to study
of Galactic kinematics using distant OB3 stars ($r<1$ kpc) with
distances estimated with accuracy $\approx15\%$ by Megier, Strobel
\& Galazutdinov (2009) from absorption lines of interstellar CaII.
Note that for majority of these OB3 stars such high-precision
estimates were obtained for the first time, taking into account
that HIPPARCOS (1997) trigonometric parallaxes for them are not
significant.

Nearby OB3 stars are of great interest for our tasks too. As can
be seen from fig.~6 from  Megier et al.~(2009), and fig.~1 from
Bobylev~\&~Bajkova~(2011), for stars with $r<1$ kpc, the distances
obtained from lines of interstellar CaII, are in good agreement
with the estimates obtained using different methods.

This work is devoted to analysis of sample of nearby stars with
distances ($r<1$ kpc from Megier et al. (2009) list for study of
kinematic peculiarities of Galactic spiral structure and the Gould
Belt stars.

\section{Data}

For OB3 stars from Megier et al.~(2009) list we took line-of -site
velocities from CRVAD-2 compilation (Kharchenko et al. 2007) and
proper motions from improved version of HIPPARCOS catalog (van
Leeuwen 2007). For spectral-double stars the comparison with data
base SB9 (Pourbaix, Tokovinin \& Batten 2004) has been done to
revise values of systemic line-of-site velocities $V_\gamma$.
Line-of site velocities were significantly corrected for a number
of stars from CRVAD-2 catalog. The whole sample consists of 258
HIPPARCOS stars with known distances, line-of-site velocities and
proper motions including both distant and nearby stars. Sample of
nearby stars ($r<1$ kpc), which is of our interest, consists of
126 stars.

  \begin{figure}[t]
 {\begin{center}
  \includegraphics[width=80mm]{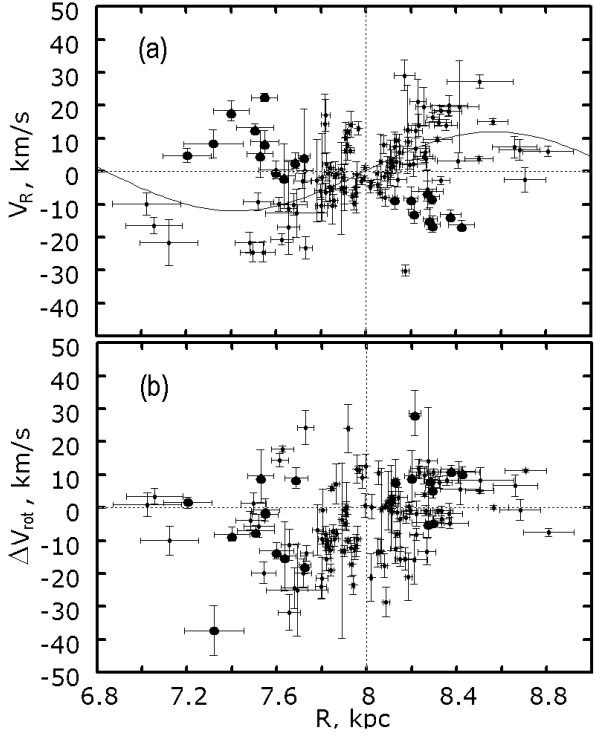}
  \caption{\small
Galactocentric radial $V_R$ velocities of 126 stars vs $R$ (a);
residual rotational velocities $\Delta V_{rot}$ (b); vertical
dotted line denotes location of the Sun.
 }
 \label{Rad-Rot}
 \end{center}}
 \end{figure}

\section{Method and basic relations}

From observations we have heliocentric line-of-sight velocity
$V_r$ in $\mathrm{km\,s}^{-1}$;  proper motion velocity components
$V_l = 4.74r\mu_l \cos b$ and $V_b = 4.74r\mu_b$ in the $l$ and
$b$ directions, respectively (the coefficient $4.74$ is the
quotient of the number of kilometers in astronomical unit by the
number of seconds in a tropical year); heliocentric distance $r$
in kpc for a star. Proper motion components $\mu_l \cos b$ and
$\mu_b$ are measured in $\mathrm{mas\, yr}^{-1}$. We adopt that
$R_0=8.0\pm0.4$~kpc is the galactocentric distance of the Sun
(Foster~\&~Cooper, 2010).

Components of spatial velocities $U,V,W$ of the stars are
determined from observed line-of-sight and tangential velocities
in the following way:
 \begin{equation}
 \begin{array}{lll}
 U&=&V_r\cos l\cos b-V_l\sin l-V_b\cos l\sin b,\\
 V&=&V_r\sin l\cos b+V_l\cos l-V_b\sin l\sin b,\\
 W&=&V_r\sin b                +V_b\cos b.
 \label{UVW}
 \end{array}
 \end{equation}
Velocity $U$ is directed towards the Galactic Centre, $V$ along
the Galactic rotation, and $W$ towards the Northern Galactic pole.

Two projections of these velocities: $V_R$, directed radially from
the Galactic Centre towards an object, and $V_{rot}$, orthogonal
to $V_R$ and directed towards Galactic rotation, are found from
the following relations:
 \begin{equation}
 \begin{array}{lll}
  V_{rot}&=& U\sin \theta+(V_0+V)\cos \theta, \\
         V_R&=&-U\cos \theta+(V_0+V)\sin \theta,
 \label{VRVT}
 \end{array}
 \end{equation}
where $V_0=|R_0\Omega_0|$, and position angle $\theta$ is
determined as $\tan\theta=y/(R_0-x)$, where $x,y$ are Galactic
Cartesian coordinates of an object. The quantity $\Omega_0$ is the
Galactic angular rotational velocity at distance $R_0$, parameters
$\Omega^1_0, \ldots, \Omega^n_0$ are derivatives of the angular
velocity from the first to the $n$-th order, respectively.

In addition, it is assumed that velocities $U$ and  $V$ are free
from the Solar velocity with respect to mean group velocity
$(U_\odot,V_\odot,W_\odot)$. According to Sch\"onrich, Binney \&
Dehnen~(2010)
 $(U_\odot,V_\odot,W_\odot)_\mathrm{LSR}=(11.1,12.2,7.3)\pm(0.7,0.5,0.4)\;\mathrm{km\, s}^{-1}$.

According to linear theory of density waves (Lin~\&~Shu 1964)
 \begin{equation}
 \begin{array}{lll}
       V_R =-f_R \cos \chi,\\
 \Delta V_{rot}= f_\theta \sin\chi,
 \label{DelVRot}
 \end{array}
 \end{equation}
where
 \begin{equation}
\chi=m[\cot(i)\ln(R/R_0)-\theta]+\chi_\odot
 \end{equation}
is phase of the spiral wave ($m$ is number of spiral arms, $i$ is
pitch angle, $\chi_\odot$ is the radial phase of the Sun in the
spiral wave; $f_R$ and $f_\theta$ are amplitudes of radial and
tangential components of the perturbed velocities which, for
convenience, are always considered positive. The distance $R$ of
an object from the Galactic rotation axis is calculated as
  \begin{equation}
 R^2=r^2\cos^2 b-2R_\circ r\cos b\cos l+R^2_\circ.
 \end{equation}

%%%%%%%%%%%%%%
 \begin{table}[t]                                     % ╥рсышЎр~1.
 \caption[]{\small HIPPARCOS numbers of the OB3 stars marked
 in fig.~\ref{Rad-Rot} by black circles. }
 \begin{center}
 \label{t:01}
 \small
 \begin{tabular}{|r|r|r|r|r|r|r|r|r|r|r|r|}\hline
 20234 &  23972 & 31088 & 32385 & 34561 & 35412 \\
 36778 &  40328 & 66925 & 80569 & 82110 & 86768 \\
 87812 &  87768 & 88331 & 88886 & 94385 & 94827 \\
 96483 & 117810 &       &       &       &   \\
 \hline
 \end{tabular}
 \end{center}
 \end{table}
%%%%%%%%%%%%%%

  \begin{figure}[t]
 {\begin{center}
  \includegraphics[width=80mm]{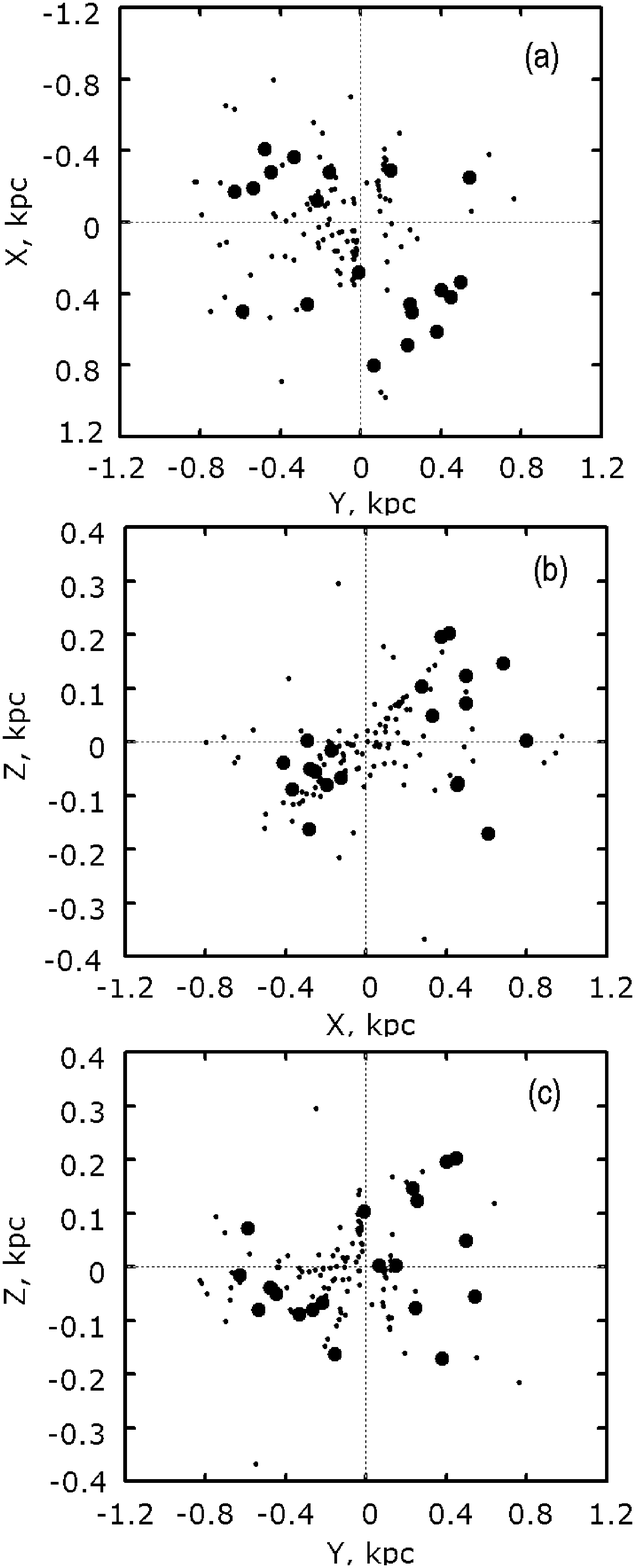}
  \caption{\small
Distribution of 126 stars in Galactic planes $XY$ (a); $XZ$(b) and
$YZ$ (c).
 }
 \label{XYZ}
 \end{center}}
 \end{figure}

In work by Bobylev~\&~Bajkova~(2011) using 102 distant OB3 stars
from Megier et al.~(2009) list, on the basis of Bottlinger
kinematic equations, adopting $R_0=8$~kpc we found components of
the solar peculiar velocity
 $$
 (u_\odot,v_\odot,w_\odot)=(8.9,10.3,6.8)\pm(0.6,1.0,0.4) {\rm~km
 s^{-1}},
 $$
galactic rotation parameters
 \begin{equation}
 \label{OMEGA}
 \begin{array}{lll}
  \Omega_0 &=& -31.5\pm0.9, ~\hbox{km s$^{-1}$ kpc$^{-1}$},\\
  \Omega^{'}_0 &=& +4.49\pm0.12,~\hbox{km s$^{-1}$ kpc$^{-2}$},\\
 \Omega^{''}_0 &=& -1.05\pm0.38,~\hbox{km s$^{-1}$ kpc$^{-3}$},
 \end{array}
 \end{equation}
amplitudes of spiral density wave
$$
 \begin{array}{lll}
 f_R=-12.5\pm1.1,~{\rm km s^{-1}},\\
 f_\theta= 2.0\pm1.6,~{\rm km s^{-1}},
  \end{array}
 $$
pitch angle for two-armed spiral structure
 $$
 i=-5.3^\circ\pm0.3^\circ,
 $$
 wavelength
 $$\lambda=2.3\pm0.2~{\rm kpc},
 $$ the radial
phase of the Sun in the spiral wave
 $$
 \chi_\odot=-91^\circ\pm4^\circ.
 $$
Having rotational parameters (\ref{OMEGA}) it is easy to construct
Galactic rotational curve and calculate residual rotational
velocities $\Delta V_{rot}.$

\section{Results}

In fig.~\ref{Rad-Rot} galactocentric radial velocities $V_R$ (a)
and residual rotational velocities $\Delta V_{rot}$ (b) are shown
for 126 OB3 stars from the Solar neighborhood $r<1$~kpc vs
galactocentric distances $R$. The harmonic curve, shown in the top
diagram, is a result of analysis of 102 distant OB3 stars (see
solution (\ref{OMEGA})), having amplitude of radial perturbations
$f_R=12.5$~km s$^{-1}$ and radial phase of the Sun in spiral
density wave $\chi_\odot=-91^\circ$. As it is seen from the
picture this harmonic curve is fitted quite well to the majority
of the stars of our sample. But there is a surprising group of 20
stars, marked by black circles, which distribution on the diagram
is strongly different. As for residual rotational velocities
(picture (b)) these stars does not show any peculiarities. So,
peculiarities are seen only in radial velocities diagram. The
Hipparcos numbers of marked unusual stars are given in table
~\ref{t:01}.

In fig.~\ref{XYZ} the distribution of our 126 stars is shown in
three galactic planes. From analysis of these diagrams we can
conclude, in the first place, that the stars of our sample are
tightly bounded with the Gould Belt structure. This conclusion
follows from a characteristic inclination  of the star system
$\approx$17$^\circ$ to Galactic plane $XY$ (Bobylev 2006), what is
apparent from the middle diagram (b). In the second place,
distribution of marked 20 stars is not distinguished from
distribution of other stars.

The analysis presented here allows us to make an important
conclusion that galactocentric radial velocities ($V_R$) of the
Gould Belt stars are quite compatible with the galactic spiral
density wave parameters found from data on distant stars.

The most interesting question is related to the nature of group of
20 marked stars. We put forth two possibilities:

1.There exist two spiral waves with almost equal amplitudes, but
shifted each other by phase $\Delta \chi_\odot\approx180^\circ$.
Similar composite models are known in literature, for example, a
model 2$+$4, supposing existence in the Solar neighborhood
simultaneously of two- and four-armed spiral structures (L\'epine,
Mishurov \& Dedikov~2001; Mishurov~\&~Acharova~2011).

2.Distances of these 20 stars are measured with large errors,
caused by non-homogeneities in distribution of interstellar CaII
clouds in the nearest Solar neighborhood connected with the Local
Bubble. Here the variation of volume density of CaII ions attains
10 times under mean volume density
$n_{CaII}\sim10^{-9}$~ёь$^{-3}$, and  size of spatial
non-homogeneities are $\approx$60~pc (Welsh et al. 2010).

We tried to eliminate the peculiarities in distribution of these
20 stars by decreasing their distances using different scaling
coefficients $Ks$: $r_{new}=r\cdot Ks$. For a number of stars
listed in table ~\ref{t:01} the trigonometric parallaxes are
determined with small errors ($\sigma_\pi/\pi<10\%$) and
corresponding trigonometric distances are about a half of
distances determined using CaII scale. Therefore firstly we
adopted $Ks=0.5$, but scaling by this coefficient did not change
significantly the character of distribution. Only if
$Ks=0.2-0.1,$ the peculiarities can be eliminated almost totally.
But such value of $Ks$ seems to be unrealistic.

\begin{figure}[t]
 {\begin{center}
  \includegraphics[width=80mm]{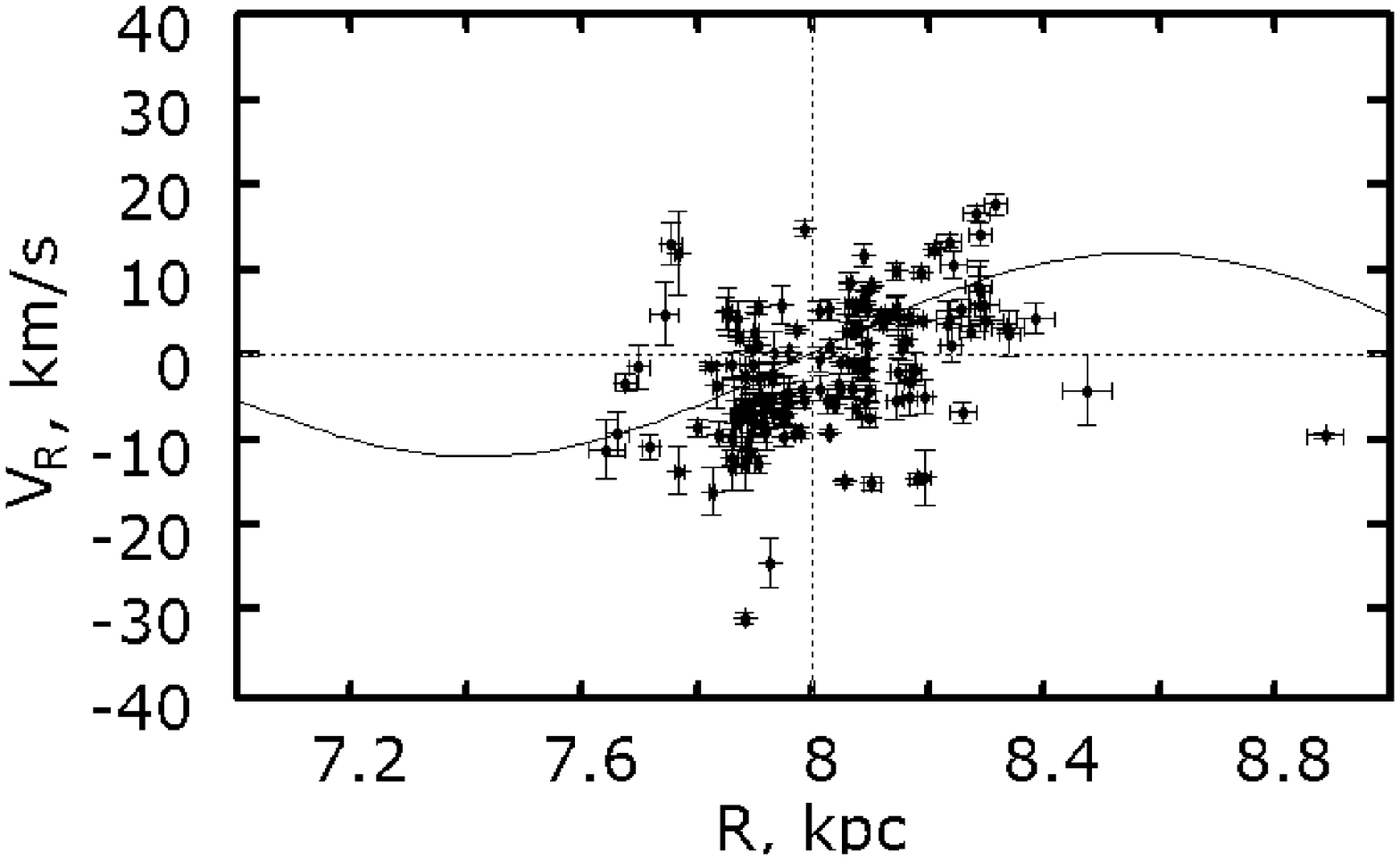}
  \caption{\small
Galactocentric radial  velocities $V_R$ of 147 HIPPARCOS stars of
spectral classes O--B2.5, which distances are determined with
errors $\sigma_\pi/\pi\leq10\%$.
 }
 \label{Rad-HIP}
 \end{center}}
 \end{figure}

The first possibility seems to be more probable. To test it we
formed a new sample of 147 HIPPARCOS stars of spectral classes
O--B2.5, which trigonometric parallaxes are determined with errors
$\sigma_\pi/\pi\leq10\%$ and line-of-site velocities from CRVAD-2
catalog (Kharchenko et al. 2007). Note that these stars have the
most reliable distances. Galactocentric radial velocities $V_R$ of
these stars are shown in fig.~\ref{Rad-HIP} together with the same
harmonic curve shown in fig.~\ref{Rad-Rot} (a). Note that in
fig.~\ref{Rad-HIP} there are a few stars from table~\ref{t:01}.
Although we can not make some definite conclusions from
fig.~\ref{Rad-HIP}, all the same we can see a weak indicator of
the presence of the second spiral pattern. Obviously, this problem
requires further investigation.

\section{Conclusions}

We tested the distances derived from the equivalent widths of
interstellar CaII spectral lines by Megier et al. To this end, we
used a sample of nearby ($r<$ 1 kpc) 126 young OB3 stars with
known proper motions and line-of-site velocities. It is shown that
these stars are tightly bounded with the Gould Belt structure.

We found that Galactic spiral density wave parameters obtained
from the galactocentric radial velocities ($V_R$) of about 100
stars of our sample are in good agreement with ones obtained
recently from sample of distant OB3 stars showing a wave with
amplitude $f_R\approx$12~km/s, wavelength $\lambda\approx2.3$~kpc
and phase of the Sun in spiral wave $\chi_\odot\approx-90^\circ$.

We revealed 20 stars with absolutely unusual kinematical features.
Their galactocentric radial velocities ($V_R$) show a wave, biased
on $\approx180^\circ$ with respect to the wave, found from the
whole sample. We proposed two hypothesis for discussion:
1)superposition of two spiral patterns; 2)errors in distances
determined from the equivalent widths of interstellar CaII
spectral lines are caused by non-homogeneity in distribution of
interstellar CaII clouds in the nearest Solar neighborhood
connected with the Local Bubble. But the first idea of
superposition of two spiral patterns seems to be the more probable
requiring further study.

\acknowledgements This work was supported by the ``Nonstationary
Phenomena in Objects of the Universe'' Program of the Presidium of
the Russian Academy of Sciences and the Program of State Support
for Leading Scientific Schools of the Russian Federation (project.
NSh--16245.2012.2, ``Multiwavelength Astrophysical Studies'').

\end{document}